\documentstyle[psfig]{laa}

\newcommand{\bib}[1]{#1}
\newcommand{\etal}{{et al.}\ }             
\newcommand{\eg}{{e.g.},\ }                
\newcommand{\cf}{{cf.}\ }                  

\newenvironment{planotable}[1]
{\begin{table}\caption[]{\ptcaption}
\begin{flushleft}\edef\tableformat{\string#1}\ptcolsep
\begin{tabular}{\tableformat}}{\noalign{\smallskip}\hline
\noalign{\medskip}\end{tabular}
\\\ptcomments\medskip\ptrefs\end{flushleft} 
\label{\ptlabel}
\end{table}}

\newenvironment{planotable*}[1]
{\begin{table*}\caption[]{\ptcaption}
\begin{flushleft}\edef\tableformat{\string#1}\ptcolsep
\begin{tabular}{\tableformat}}{\noalign{\smallskip}\hline
\noalign{\medskip}\end{tabular}
\\\ptcomments\medskip\ptrefs\end{flushleft}
\label{\ptlabel}
\end{table*}}

\def\tablecolsep#1{\gdef\ptcolsep{\tabcolsep=#1}} \def\ptcolsep{\relax}
\def\tablecaption#1{\gdef\ptcaption{#1}} \def\ptcaption{\relax}
\def\tablelabel#1{\gdef\ptlabel{#1}} \def\ptlabel{\relax}

 \def\ptcomments{\relax}

 \def\ptrefs{\relax}

\newcommand{\nl}{\\}
\newcommand{\tablehead}[1]{\hline\noalign{\smallskip}#1\\}
\newcommand{\colhead}[1]{\multicolumn{1}{c}{#1}}
\newcommand{\startdata}{\noalign{\smallskip}\hline\noalign{\smallskip}}

\newcommand{\tablewidth}[1]{\typeout
{----- tablewidth not implemented with A\&A ----------------------}}{}
{}

\begin{document}

\thesaurus{11.06.2; 11.09.1 Tucana; 11.12.1; 11.19.5; 11.19.6}

\title{ CCD photometry of the Tucana dwarf galaxy
\thanks{Based on data collected at the European Southern Observatory,
         La Silla, Chile}
}
\subtitle{}

\author {Ivo Saviane \inst{1} \and Enrico V. Held \inst{2}
\and Giampaolo Piotto\inst{1} }

\offprints {E. V. Held}

\institute{
Universit\`a di Padova, Dipartimento di Astronomia, vicolo dell'Osservatorio 5, 
I--35122 Padova, Italy
\and
Osservatorio Astronomico di Bologna, via Zamboni 33, I--40126 Bologna, Italy
}

\date {}

\maketitle

\begin{abstract}
We present $V$ and $I$ CCD photometry for $\sim360$ stars 
in the recently discovered dwarf galaxy Tucana. 
The large field investigated and the accurate photometric calibration
make our data complementary to the deeper HST photometry.
From the $I$ magnitude of the tip of the red
giant branch we estimate a distance modulus $(m-M)_0=24.69\pm0.16$,
corresponding to $870\pm60$ Kpc, confirming that Tucana is an isolated
dwarf spheroidal located almost at the border of the Local Group. 
From the color of the red giant branch tip and by 
direct comparison with the
giant branches of galactic globular clusters we estimate a
metallicity [Fe/H]=$-1.8\pm0.2$, with no clear indication for a metallicity
spread. 
The color--magnitude diagram indicates that Tucana has had a single star 
formation burst at the epoch of the Galactic globular cluster star 
formation. There is no evidence for an intermediate or young stellar 
population. 
We derive the $V$ luminosity profile, the surface density profile 
of resolved stars, and the structural parameters of Tucana,
from which we confirm that 
Tucana participates to the general 
metallicity--surface brightness--absolute magnitude relations
defined by the 
Galaxy and M31 dwarf spheroidal and dwarf elliptical companions.


\end{abstract}

\section{Introduction}
\label{sec_intro}

Among the 12  known dwarf spheroidal (dSph) galaxies in the Local Group 
(LG), nine
(including the recently discovered
Sagittarius dwarf; \bib{Ibata \etal 1994})
are in the halo of the Milky Way, three are companions to the Andromeda
Galaxy, and only one, Tucana, is
far from any luminous member of the LG.
The properties of the LG
are the subject of recent reviews, \eg \bib{Caldwell (1995)}, 
\bib{Da Costa (1994)}, \bib{Gallagher \& Wyse (1994)}, and \bib{Zinn (1993)}.
The isolated location of Tucana offers a unique  opportunity to
study the evolution of dwarf galaxies in an environment different from the
halo of  giant spiral galaxies and to constrain models for their formation.

Tucana, although already present in the Southern Galaxy
Catalog (\bib{Corwin \etal 1985}), was overlooked for some time before its
serendipitous re-discovery by \bib{Lavery (1990)}.
\bib{Lavery \& Mighell (1992)} presented a shallow
$V$ -- $(V-I)$ color-magnitude
diagram (CMD)  based on low-resolution imaging reaching $V \sim 23$.
They suggested the classification of this galaxy as a dE5 dwarf spheroidal in
the LG, and an upper limit of 24.75 mag for the distance
modulus. No direct metallicity determination was given.

Preliminary photometric results have been reported by 
\bib{Da Costa's (1994)} in his recent review. 
He obtained a $V$ -- $(B-V)$ color-magnitude 
diagram showing neither young blue stars nor very red, 
bright stars that might indicate the
presence of a conspicuous intermediate-age population, from which
Tucana seems to be 
indistinguishable from the dSph companions to our Galaxy and M31. 
By fitting the giant branches of M92, M3, and 47~Tuc to the data, he obtained
$(m-M)_0 = 24.8 \pm 0.2$ mag
and a metallicity [Fe/H] $= -1.8 \pm 0.25$ dex,
the distance error being the largest contributor to the uncertainty in the
abundance.
\bib{Da Costa (1994)} also presented a $B$ surface brightness profile of Tucana
showing no indication for sub-structure in the star
distribution.
An exponential law fit to the data yields a scale length $\alpha =
30$\arcsec\ (geometric mean radius) or 130 pc at a distance of 900 kpc,
an extrapolated central surface brightness
$\mu_{0,B} = 25.5$ mag arcsec$^{-2}$, and
an absolute magnitude M$_V = -9.3 \pm 0.4$.

The first results of an ongoing deep HST imaging study of
Tucana by Seitzer and coll. (see \bib{Caldwell 1995}) are
a color-magnitude diagram with no evidence for
young stars, and a moderately blue horizontal branch (HB).
Since Tucana has a metallicity as low as [Fe/H] $= -1.8$
(a value confirmed by this study), one
might expect an even bluer HB, and the authors suggest that
a mild second parameter effect may be present
 (\cf \eg Leo~II: \bib{Demers S. \& Irwin 1993}).
\bib{Castellani M. \etal (1995)} presented a progress report of a study
using the same data (obtained from the HST archive),
and deep ground-based CCD observations. They obtain a distance
modulus consistent with that of \bib{Da Costa (1994)}, and marginally higher
metallicity ([Fe/H] $= -1.56 \pm 0.20$), and claim that the spread in
color on the RGB is larger than the one expected on the basis of photometric
errors.

In order to provide clues as to whether the star formation history and
structure of dwarf spheroidals  are affected by environment, we present 
a comprehensive CCD study of the stellar populations and structure of Tucana,
based on $V,I$ CCD photometry over a relatively wide field of $5\farcm7 \times
5\farcm7$. This area allows a complete coverage of the galaxy.
Photometry of the stars in our field provides 
a well populated upper RGB, essential for 
obtaining reliable distance and metallicity estimates. 
The plan of the present paper is as follows. 
Observations and stellar photometry are described in Sec.~\ref{sec_obsred}.
Sect.~\ref{sec_cmd}  presents
the $I$ -- $(V-I)$  color-magnitude diagrams of Tucana. 
A new accurate distance determination is
derived in Sec.~\ref{sec_lf} 
by locating the RGB tip on the $I$ luminosity function.
New estimates for the mean metal abundance and abundance spread of
stars on the red giant branch (RGB) are presented in Sec.~\ref{sec_metal}. 
Surface photometry is presented in  Sec.~\ref{sec_surfot}
where structural and photometric parameters are derived.
In Sec.~\ref{sec_discu} we discuss the properties of Tucana in the framework
of the general relations for dwarf elliptical and spheroidal galaxies. 

\section{Observations and data reduction}
\label{sec_obsred}

\begin{figure}
\psfig{figure=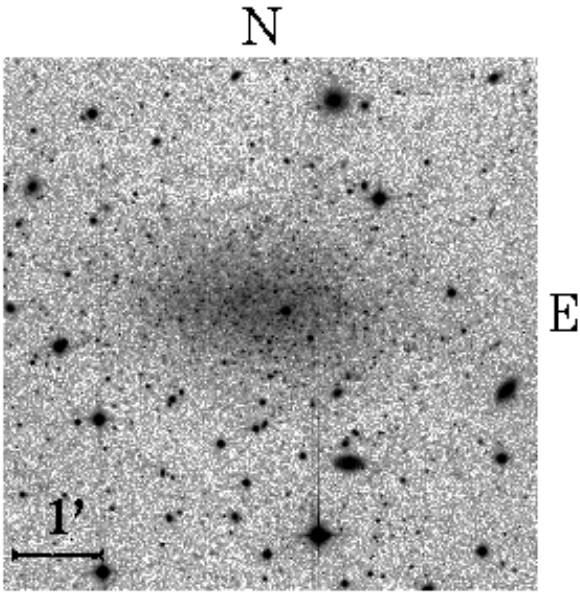,width=8.5cm}
\caption[]{The combined V image of Tucana from our data with good seeing}
\label{v_sum}
\end{figure}

Observations were obtained in two runs, on September 15-16, 1993 and
September 2--3, 1994, using EFOSC2 at the ESO 2.2m telescope.
EFOSC2 was equipped with a 1024 $\times $ 1024 pixels Thomson coated
CCD (ESO \#19); the pixel size was 19 $\mu$m yielding a scale 0\farcs336
pixel$^{-1}$ and a field of view 5\farcm7 $\times $ 5\farcm7.
The CCD conversion factor was 2.1 e$^{-}$/ADU, and the read-out noise,
measured
from the overscan region,  was 4 e$^{-}$ (\cf EFOSC2 User Manual).

Weather conditions and seeing were quite variable during both runs, so
the quality of the observations is not uniform.
A journal of the observations selected for this study is given in
Table~\ref{obsjou}.
The useful images have been grouped into two sets:
the first set includes the best images
(FWHM $\leq$ 1.2\arcsec), most of which were taken on the single best
night of Sept. 3, 1994 (under photometric conditions).
A second set includes all the images taken
in medium, yet acceptable, seeing conditions ($\sim$ 1\farcs5  FWHM).
All images with seeing worse than $1.5\arcsec$ FWHM have been discarded.

\tablecaption{The journal of observations }
\tablelabel{obsjou}
\begin{planotable}{cclllcc}
\tablehead{
\colhead{$N$} &
\colhead{filter} &
\colhead{t$_{\rm exp}$} &
\colhead{Date} &
\colhead{UT} &
\colhead{$X$} &
\colhead{FWHM} }
\startdata
 1   &  V & 1800   &  1993~Sept.~15  & 01:40  & 1.32 &  1.4\arcsec \nl
 2   &  I & 2700   &  1993~Sept.~15  & 02:15  & 1.27 &  1.5\arcsec \nl
 3  &  V & 1800    &  1993~Sept.~15  & 03:09  & 1.23 &  1.3\arcsec \nl
 4  &  I & 2700    &  1993~Sept.~15  & 03:44  & 1.22 &  1.5\arcsec \nl
 5  &  V  & 1800  &   1994~Sept.~3  &  00:58  & 1.52   & 1.2\arcsec  \nl
 6  &  I  & 1800  &   1994~Sept.~3  &  01:33  & 1.42   & 1.2\arcsec  \nl
 7  &  I  & 1800  &   1994~Sept.~3  &  02:06  & 1.35   & 1.5\arcsec  \nl
 8  &  V  & 1800  &   1994~Sept.~3  &  02:46  & 1.29   & 1.5\arcsec  \nl
 9  &  V  & 1800  &   1994~Sept.~3  &  03:19  & 1.26   & 1.5\arcsec  \nl
 10  &  I  & 1800  &  1994~Sept.~3  &  03:55  & 1.23   & 1.2\arcsec  \nl
 11  &  I  & 1800  &  1994~Sept.~3  &  04:29  & 1.22   & 1.1\arcsec  \nl
 12  &  I  & 1800  &  1994~Sept.~3  &  05:02  & 1.23   & 1.2\arcsec  \nl
 13  &  V  & 1800  &  1994~Sept.~3  &  05:42  & 1.25   & 1.2\arcsec  \nl
\end{planotable}

Processing of the CCD frames was accomplished as follows.
First, a constant offset was determined form the overscan region and
subtracted from each frame (sample bias frames are constant across the chip
and stable in time).
Both sky and dome flat-fields
(the latter taken on a white screen illuminated by day light)
were taken in the 1994 run.
A comparison of dome and sky flats shows them to be consistent at
the 1\% level.
For the 1994 observations, master twilight-sky flats have been
created for each filter, with
residual stellar images removed by taking the median of individual exposures.
For the 1993 run only dome flats were available.

After preliminary processing, the frames were
sorted by filter and by seeing quality.
Images from the ``good seeing'' set and the  ``medium seeing'' set
were carefully registered to a common reference frame.
Finally, the registered frames were co-added, yielding four master
images, and the 1.2\arcsec\ $V$ image is shown in Fig.~\ref{v_sum}.
The size of the Point Spread Function (PSF) 
on the combined images shows no degradation with respect to
the individual frames.

Stellar photometry was performed using DAOPHOT and ALLSTAR
(\bib{Stetson 1987, 1994}) in the MIDAS environment.
After a few experiments, we found that an analytic Moffat function (with
$\beta$ = 2.5), along with a variable PSF model
with quadratic dependence on distance
from the frame center, provided the best representation of the PSF.
A PSF was iteratively constructed for each master frame
starting  with a set of $\sim 50$ bright, non-saturated stars and
progressively reducing this sample to $\sim$20 stars
by discarding poorly fit stars.
In particular, visual inspection of a PSF image created by adding an artificial
star to an empty frame was employed to identify and reject any PSF stars
contaminated by faint components.
ALLSTAR was run twice on the combined frames, using our best PSF and the
aperture photometry catalogs. For each of the four combined
images, all stars detected and analyzed on the first run of ALLSTAR were
subtracted and the residual images searched again for faint undetected objects.
Then, ALLSTAR was run again on the complete list of stars.

In order to analyze the degree of completeness and the photometric
errors as a function of magnitude and location on the frame,
artificial star experiments were performed using the DAOPHOT
task ADDSTAR, along with
a code written by one of us (I.S.) to perform automatically the individual
experiments as well as the general program of simulations.
Each experiment follows a scheme designed to yield approximately
constant number of stars per magnitude bin.
The results of these simulations are summarized in Table~\ref{c_and_err}:
columns 1 and 2 specify the central magnitude of the $V$ bins and mean
colors, columns 3 to 5 are the standard deviations in $V$, $I$, and $(V-I)$,
column 6 to 8 give
the number of stars added ($N_{\rm i}$) for each magnitude bin,
the completeness ratio  $c = N_{\rm f}$/$N_{\rm i}$ 
(where $N_{\rm f}$ is the number of artificial stars eventually found and
measured by DAOPHOT), and its standard error.
Magnitude (color) errors were derived from the rms scatter of the
difference between input and measured magnitudes (colors) of the artificial
stars.
Finally, $\sigma_{c}$ was estimated from the scatter of $c$ obtained from
10 individual experiments at a fixed magnitude. Completeness is $\sim1$ for
$V \leq$ 20.7 and falls below 20\% for $V \geq$ 24.2.

\tablecolsep{4pt}
\tablecaption{Photometric errors and incompleteness}
\tablelabel{c_and_err}
\begin{planotable}{cclllrcl}
\tablehead{
\colhead{$V$} &
\colhead{$(V-I)$} &
\colhead{$\sigma_{V}$} &
\colhead{$\sigma_{I}$} &
\colhead{$\sigma_{(V-I)}$} &
\colhead{$N_{\rm i}$} &
\colhead{$N_{\rm f}$/$N_{\rm i}$} &
\colhead{$\sigma_{c}$}
}
\startdata
 21.24 &  1.60 &  0.014 &  0.010 & 0.021 & 300  & 0.99  &  0.03 \\
 21.74 &  1.51 &  0.021 &  0.017 & 0.030 & 420  & 0.99  &  0.02 \\
 22.23 &  1.42 &  0.028 &  0.028 & 0.045 & 800  & 0.99  &  0.03 \\
 22.73 &  1.33 &  0.041 &  0.044 & 0.071 & 1240 & 0.98  &  0.03 \\
 23.22 &  1.21 &  0.071 &  0.068 & 0.093 & 2060 & 0.95  &  0.03 \\
 23.71 &  1.09 &  0.108 &  0.109 & 0.153 & 2500 & 0.84  &  0.04 \\
 23.96 &  1.02 &  0.110 &  0.159 & 0.193 & 220  & 0.64  &  0.09 \\
\end{planotable}

Star-like objects were selected out of the  raw photometric catalogs
by using a set of selection criteria based on the values of
the DAOPHOT parameters related
to the quality of the fit and image shape ({$\chi$} and {\em  sharp}).
The distribution of the same parameters  resulting from artificial
star experiments served as a training set, complemented by careful scrutiny of
the Tucana co-added images.
Objects with $\chi > 1.5$  turned out to
be hot pixels (mostly cosmic rays) or spikes around bright sources; similarly,
objects with {\em  sharp} $< -1$ are isolated hot pixels while
{\em  sharp} $> 1$ corresponds to PSF spikes and diffuse objects.

Absolute calibration is based on observations
of \bib{Landolt's (1992)} standard stars on the photometric
night of Sept.~3, 1994.
We have obtained the following calibrations
(standard deviations of the residuals are 0.006 mag in $V$
and 0.014 mag in $I$, respectively):

\bigskip
\noindent
\begin{eqnarray}
 V & = & {v^\prime} + 0.053\,(V-I) +  23.69 \\
 I & = & {i^\prime} - 0.055\,(V-I) +  22.47
\end{eqnarray}
\bigskip

\noindent
where $v^\prime$, $i^\prime$ are
instrumental magnitudes defined as follows:

\begin{equation}
m^\prime = m_{\rm ap} + 2.5\log(t_{\rm exp} + \Delta\,t) -
k_{\lambda} X.
\end{equation}

\noindent
$m_{\rm ap}$ are the ``total'' magnitudes resulting from
a growth-curve analysis of the
standard stars in circular apertures up to a radius $R=6\farcs4$
(we have verified that magnitudes measured within larger apertures, 
say $R\le 10\arcsec$, are only 0.01 mag brighter on the average).
$\Delta\,t$ is the average shutter delay
and $k_{\lambda}$ represents the extinction coefficients in the $V, I$ bands.
For the shutter delay we have assumed 0.4 s (Veronesi 1995); 
assuming $\Delta\,t = 0$ would
not affect the $(V-I)$ color measurements, while
$V$, $I$ magnitudes would be fainter by 0.02 mag.
We used mean extinction coefficients for La Silla
$k_{V}$ = 0.14  and $k_{I}$ = 0.06, obtained from photoelectric photometry
on the standard Johnson/Bessell system at La Silla
(G. Clementini, priv. comm.). These values are confirmed by the La Silla
Extinction database of the Geneva Observatory.
A plausible $\pm 10$\% fluctuation of extinction during the night
may contribute about $\pm 0.02$\ mag to the zero point uncertainty on the
$V$ and $I$ magnitudes, and less than 0.005 mag to the  
$(V-I)$ color uncertainty.

The aperture corrections needed to transform the PSF magnitudes of the Tucana
stars into ``total'' aperture magnitudes were then calculated as mean
differences ($C_V$ and $C_I$) between raw PSF-fit magnitudes and
the instrumental magnitudes, for the set of bright, isolated stars on  
reference images of Tucana taken on the photometric night of Sept. 3, 1994.
We paid particular attention to avoid systematic errors on the colors, and
found that the instrumental colors of the reference stars are 
almost independent of aperture size (with an rms error of
0.004 mag) for radii $3\farcs0 < R < 6\farcs4$
(magnitudes measured in larger apertures become more affected by 
background noise).
To assess the systematic accuracy of the
calibrated magnitudes and colors, we  compared the instrumental magnitudes
$v^\prime$, $i^\prime$ of the reference stars
from {\em all} the individual images of Tucana taken on the
photometric night of Sept. 3, 1994.
Table~\ref{gamma} lists the values of the zero-point constants
$C_V$ and $C_I$ for all the individual calibration frames.
The $C_V$, $C_I$ values show some scatter (0.01 mag in $V$ and 0.02 mag
in $I$, $1\sigma$ errors), but no obvious dependence on seeing.
This scatter includes, in addition to uncertainties on aperture correction,
the errors on shutter timing and extinction corrections.
These errors, together with the uncertainty on the calibration
relations, yield a 0.01 mag zero-point rms uncertainty in the $V$ band,
and 0.02 mag in the $I$ band. The corresponding uncertainty on the
$(V-I)$ color is $0.03$\  mag.

\tablecaption{Zero point shifts as determined from different images}
\tablelabel{gamma}
\begin{planotable}{rrccc}
\tablehead{
\colhead{$N$} &
\colhead{filter} &
\colhead{FWHM} &
\colhead{airmass} &
\colhead{$C_{\lambda}$}
}
\startdata
  6      & I &   1.2 & 1.42  &  7.56 \nl
  7      & I &   1.5 & 1.35  &  7.57 \nl
  10     & I &   1.2 & 1.23  &  7.63 \nl
  11     & I &   1.1 & 1.22  &  7.63 \nl
  12     & I &   1.2 & 1.23  &  7.64 \nl
  5      & V &   1.2 & 1.52  &  7.48 \nl
  8      & V &   1.5 & 1.29  &  7.50 \nl
  9      & V &   1.5 & 1.26  &  7.48 \nl
  13     & V &   1.2 & 1.25  &  7.53 \nl
\end{planotable}

\section{Color-magnitude diagrams }
\label{sec_cmd}

In order to reduce the effects of field contamination on the color-magnitude
diagrams and luminosity functions, we defined an ``inner region''
(or ``galaxy region'') including all stars within an ellipse (area 5.8
arcmin$^2$) with semi-major axis $a = 108\arcsec$ and a provisional axial
ratio $b/a = 0.56$ (adopting the axial ratio determined in
Sec.~\ref{sec_surfot} would not change our results).
Similarly, we defined an ``outer region'' of our frame (area 20.8 arcmin$^2$)
external to an ellipse with $a = 161\arcsec$.

\begin{figure}
\psfig{figure=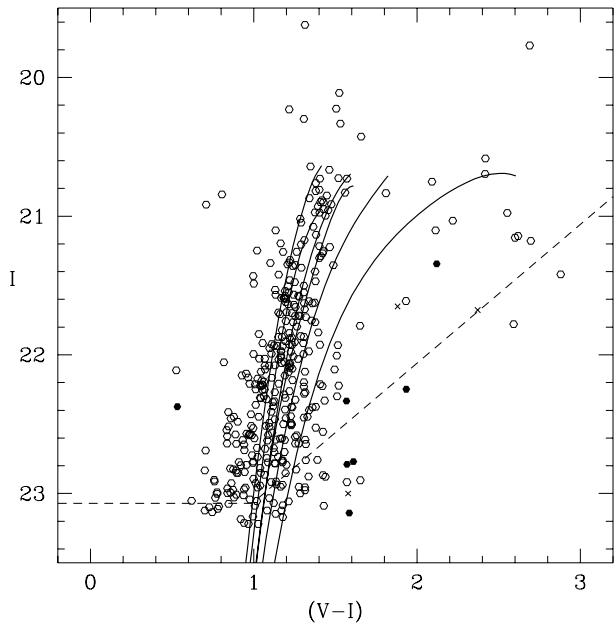,width=8.8cm}
\caption{
The Tucana's CMD, compared with globular
cluster giant branches from Da Costa \& Armandroff (1990).
Open circles represent starlike objects; obvious faint galaxies are indicated
by filled symbols, while crosses identify photometric blends; the {\it dashed
line} represents the 50~\% completeness level.
Globular cluster fiducial loci are shifted to the
same distance modulus as Tucana, assuming
$(m-M) = 24.69$ and $A_{I} = E_{(V-I)} = 0$.
 Left to right: M15 ([Fe/H]$= -2.17$), NGC~6397 ([Fe/H]=$ -1.91$),
M2 ([Fe/H]=$ -1.58$), NGC~6752 ([Fe/H]=$ -1.54$), NGC~1851
([Fe/H]=$ -1.29$) and 47~Tuc ([Fe/H]=$ -0.71$)
}
\label{cmdi}
\end{figure}

\begin{figure}
\psfig{figure=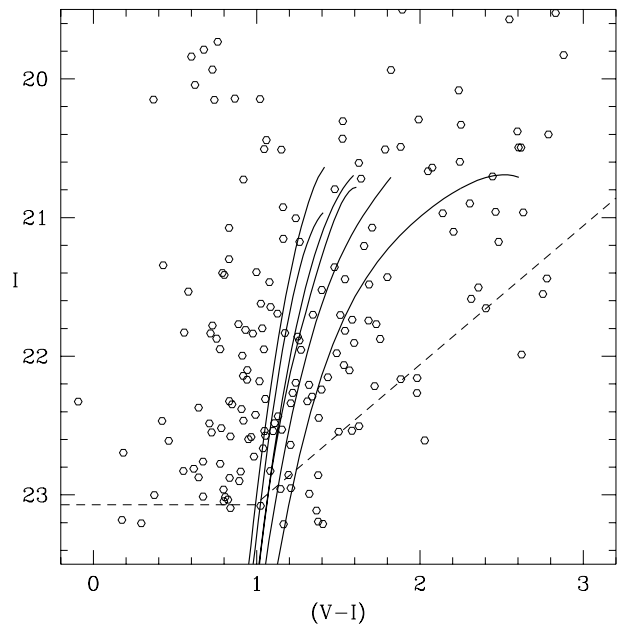,width=8.8cm}
\caption{As for Fig.~2,  for the outer field }
\label{fieldi}
\end{figure}

The color-magnitude diagram of Tucana is shown in Fig.~\ref{cmdi}, 
where the $I$ magnitude is  plotted
against the $V-I$ color; Fig.~\ref{fieldi} shows the $I$ -- $(V-I)$
color-magnitude  diagram for the stars in the outer region.
The  red giant branch of Tucana is similar to the giant branch
of old, metal-poor Galactic globular clusters.
In the same figures, the Tucana's CMD is compared with
 the $M_I$ -- $(V-I)_0$ diagrams of 6 globular clusters from Table~10
of \bib{Da Costa \& Armandroff (1990; hereafter DA90)}, suitably
scaled to the distance of Tucana.
The average metal abundance of Tucana is clearly in the range
$-2.17 <$ [Fe/H]$< -1.54$.

The color-magnitude diagram of Fig.~\ref{fieldi} is fairly
representative of the background and foreground contamination,
although Tucana extends to the outer region.
The surface density $\Sigma$\
of objects in the outer region is $\sim14$\% that in the
galaxy region (65.2 and 9.0 objects/arcmin$^2$, respectively).
The  expected ratio, as estimated for an exponential luminosity profile
with scale length $30\arcsec\pm3\arcsec$ (see Sect.~\ref{sec_surfot}),
is $\Sigma_{\rm out}/\Sigma_{\rm in} = 0.024\pm0.008$\ (the error is obtained
from the scale length uncertainty).
From this ratio and the object counts, it is easy to estimate the
contributions of galaxy stars and foreground/background
objects  in each region. It turns out
that the surface density of background objects is $\Sigma_{\rm bkg}=7.6$
objects/arcmin$^2$, i.e.
84\% of the objects in the outer region are due to background/foreground,
while the contamination of the inner sample is at the 12\% level.

An estimate of the expected Galactic star density towards Tucana
was obtained from the \bib{Bahcall \& Soneira's (1980)} model (see also
\bib{Bahcall 1986} and references therein). The ``export code'' was
modified to have $(V-I)$ as the output color instead of the original $(B-V)$.
The main modification was the replacement of the globular and open cluster
fiducial loci in $V$ -- $(B-V)$ with the corresponding $I$ -- $(V-I)$ loci.
These loci have been obtained as follows: (a) the Population I 
Main Sequence (MS) from 
\bib{Girardi et al. (1996)}; (b) the disk Population MS and the M67 MS 
from \bib{Carraro \etal (1994)}; and 
(c)  the GC's sequences from
\bib{Bergbush \& VandenBerg (1992)}
 isochrones, where the $(V-I)$  colors have been
calculated by means of \bib{Kurucz (1991)} model atmospheres.
The   $A_V/E_{(V-I)}$ ratio was adopted from
\bib{Savage \& Mathis (1979)}.
In the outer region, the color and magnitude distributions
are consistent with model expectations above the RGB tip ( $V \simeq 22$),
while there is a significant overabundance of objects with respect to the model
counts for $V > 22$ (by a factor of 2--3).

However, in the magnitude interval shown in Fig.~\ref{cmdi} some
contamination of background galaxies 
is expected.
According to \bib{Tyson's (1988)} galaxy counts, we expect
$\sim90$ objects for $20 < I < 22$. In the outer region, there are 86 
stars in this range.
According to the previous discussion, 72 of these should be 
foreground/background objects. In the same area and magnitude interval, 
we expect 26 foreground stars on the basis of the Bahcall and Soneira's 
(1980) model. Therefore, about 46 objects should be background 
unresolved galaxies.
This number is smaller than
expected by Tyson's (1988) counts; on the other hand, 
both the DAOPHOT task FIND and our rejection criteria 
tend to reject 
fuzzy not--star--like objects, and it is plausible that a fraction of the 
background galaxies has been rejected by the PSF--fitting routine, as 
demonstrated by visual inspection of the star--subtracted image.

A fiducial line through the RGB was then derived by analyzing
the color distributions in different magnitude bins.
The magnitude range of each bin, the modal $(V-I)$ color, and the color
dispersion $\sigma_{(V-I)}$, are given in Table~\ref{fiducial}
for both the 1\farcs2 images and the 1\farcs5 images.
Column 5 will be described in detail in Sec.~\ref{sec_metal}.
The modal color represents the mode of the $(V-I)$ distribution, after all the
outlier stars have been removed by an interactive procedure equivalent
to a $3\sigma$ clipping.
Trend of the color spread as a function of magnitude was smoothed
by fitting a linear relation to $\sigma_{(V-I)}$ vs. $I$:
$\sigma_{(V-I)} = 0.058\,I - 1.087$.

Stars within $2\sigma$ from the ridge line of the Tucana's RGB define
a clean sample of stars most likely belonging to Tucana
(hereafter referred to as ``$2\sigma$'' sample).
For the ``$2\sigma$'' sample, there are 339 and 87 objects in the
inner and outer region, or $\Sigma_{out}/\Sigma_{in} = 0.07$, indicating that 
there is some contamination due to foreground/background objects, but that this 
is less than 5\% 
(cf. previous discussion).

\tablecaption{Fiducial sequence along the RGB}
\tablelabel{fiducial}
\begin{planotable}{lcccc}
\tablehead{
\colhead{$I$} &
\colhead{$(V-I)$} &
\colhead{$\sigma_{(V-I)}$(1\farcs2)} &
\colhead{$\sigma_{(V-I)}$(1\farcs5)} &
\colhead{$\sigma_{\rm meas}$}
}
\startdata
 21.0--22.0  &  1.22 &  0.10 & 0.13 & 0.10 \\
 22.0--22.5  &  1.16 &  0.14 & 0.11 & 0.13 \\
 22.5--23.0  &  1.06 &  0.18 & 0.18 & 0.11 \\
\end{planotable}

We note a few starlike objects in the CMD at $(V-I) > 2$, just
above the $V$ completeness limit.
Visual inspection of these stars allows us to rule out
that they are artifacts near the limit of our photometry.
Although there are a few obvious faint galaxies (filled symbols in
Fig.~\ref{cmdi}), or photometric blends (crosses),
most of the red objects appear to be star-like.
It is of interest to understand whether these objects are red stars belonging
to Tucana, since in that case they might indicate the presence of a metal-rich
component.
There are 15 red ($V-I>2.0$) objects in the inner region and 35 in the outer 
region.
Assuming for Tucana an exponential profile implies an excess of $5.3\pm4.2$
objects in the inner region, which is significant only at the 1$\sigma$ level.
We conclude that there is no significant
population of red objects in Tucana.

As already noted by Da Costa (1994),
another feature in the CMD of Fig.~\ref{cmdi}
is the absence of any significant number
of stars bluer than the RGB.
\bib{Lavery \& Mighell (1992)} noted a concentration of stars
at $(V-I) \sim 1$, $V \sim 22$\  in their color-magnitude diagram. These may be
part of an asymptotic giant branch or, as the authors point out,
just a manifestation of their large photometric color errors.
Our CM diagrams confirm
the latter explanation: there is no excess of stars bluer than
$(V-I)=1$ at the level of the RGB tip.
More quantitatively, in the inner region
there are 8 stars with $22 < V < 22.6$\ and $(V-I) < 1.2$, while 14 stars are
found in the outer region within the same magnitude and color ranges.
Taking into account the expected surface density ratio of Tucana stars, we find
that 6 out of the 8 stars are due to foreground/background contamination, and
conclude that the number of stars bluer than $V-I = 1.2$ in Fig.~\ref{cmdi}
is consistent with contamination alone.

Bright AGB stars belonging to an intermediate-age stellar
population are quite common in dSph's
(\eg \bib{Freedman 1994}). In order to set
an upper limit to a younger stellar component in Tucana, we have estimated
the number of stars brighter than the RGB tip in excess over the field
contamination.
The bolometric absolute magnitude of the brightest AGB stars 
is $M_{\rm bol}=-3.8$ and  $M_{\rm
bol}=-4.6$ for a 10 Gyr and 3 Gyr population, respectively, 
according to the relation between AGB tip luminosity and age
given in \bib{Mould \& Da Costa (1988)} 
(\cf also the models of \bib{Bressan \etal 1994}). 
These correspond, for the Tucana's metallicity, to $M_I=-4.3$\ and
$M_I=-5.1$.  Therefore, adopting a distance modulus $(m-M) = 24.7$,
AGB stars older than 3 Gyr and younger than 10 Gyr would be found
in the range $19.6 < I < 20.4$.  Counting stars in this magnitude range
for the Tucana's inner and outer samples, we have found
7 stars in the inner region and 16 in the outer region (corresponding to
4.5 when scaled to inner area). Thus in the galaxy region there is an excess of
$2.5\pm2.9$ stars  brighter than the RGB tip ($1\sigma$\ Poisson error).
If we limit ourselves to stars redder than the RGB tip ($V-I > 1.4$)
the numbers are 4 and 7 stars, respectively, yielding an excess of
$2.0\pm2.1$ stars.
Visual inspection of the objects brighter than the RGB tip in the galaxy
region shows that at least one has an elongated shape suggesting it is a
blend or a background galaxy.
Thus we conclude that the surface density of stars brighter than
the RGB tip is entirely consistent with the star counts in the outer
region over the same magnitude range, and rule out any significant
population of AGB stars younger than 10 Gyr in Tucana.

\section{Luminosity function and distance}
\label{sec_lf}

\begin{figure}
\psfig{figure=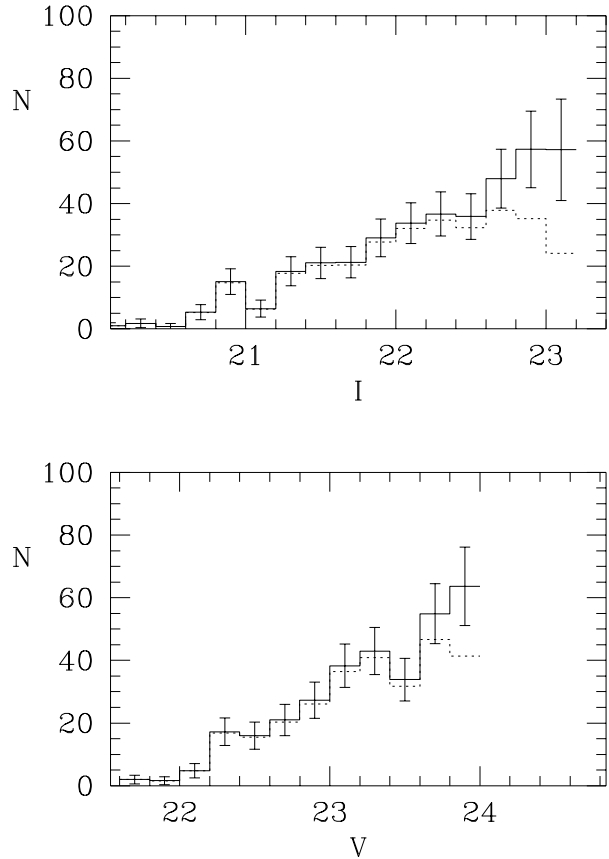,width=12.0cm}
\caption{$V$ and $I$ luminosity functions before (dotted lines) and after (solid
lines) completeness corrections and field object subtraction}
\label{lf}
\end{figure}

Assuming that Tucana does not contain very red stars, we have derived the $V$\
and $I$\ RGB luminosity functions (LF) using stars in the $2\sigma$\  sample,
with a correction for incompleteness in both $V$ and $I$ according to the
results of artificial star experiments and a correction for background/
foreground objects. The field object counts (properly scaled to the inner area)
were obtained by counting 
stars in the outer field, applying the same color selection as for the inner 
field, and allowing for the 16\% of the stars which represent 
the estimated contribution of 
Tucana stars in the outer region (cf. Sect.~\ref{sec_cmd}). 
The corrected LF's are shown if Fig.~\ref{lf}, and the corresponding tables
are available from the authors.

\begin{figure}
\psfig{figure=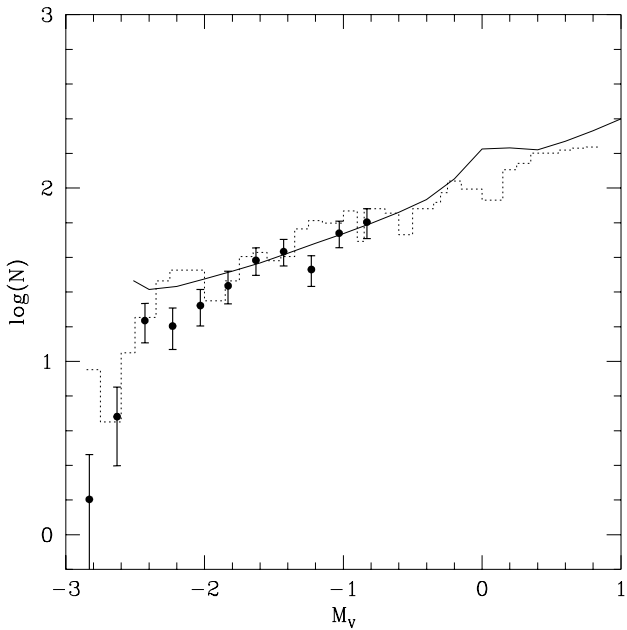,width=8.8cm}
\caption{Tucana's LF (dots with error bars)
compared to that of Fornax
(dotted histogram; Sagar et al. 1990),
and to a theoretical LF (solid line; models from Bergbush \& VandenBerg 1992,
with $Z=0.0004$ and Age = 15~Gyr)
}
\label{compare_lf}
\end{figure}

\begin{figure}
\psfig{figure=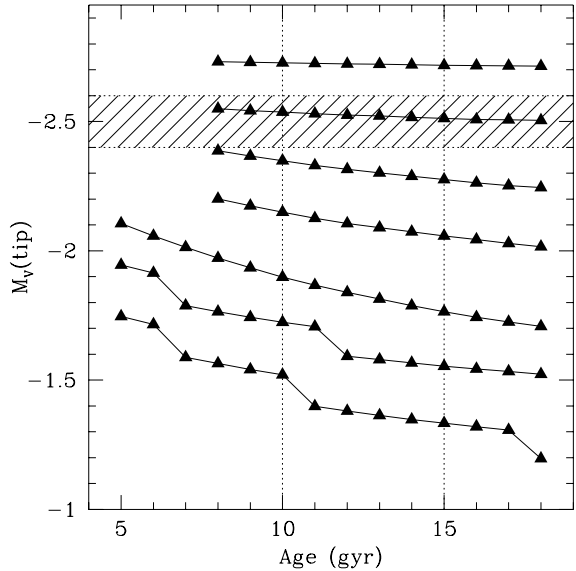,width=8.8cm}
\caption{V absolute luminosity of the RGB tip in the
 \bib{Bergbush \& VandenBerg (1992)} model isochrones, for different ages
 and metallicities.
 The shaded region indicates the absolute magnitude
 and uncertainty of the Tucana's RGB tip, after correction for distance modulus.
 Data points connected by solid lines represent models with fixed metallicity
 (top to bottom: $Z=0.0001, 0.0004, 0.001, 0.002, 0.003, 0.004, 0.006$)
 }
\label{tips_age}
\end{figure}

Figure~\ref{compare_lf} shows the RGB luminosity function of
Tucana along with the LF of Fornax (\bib{Sagar et al. 1990}).
In the same figure, the LF of Tucana is also compared with the theoretical LF
from \bib{Bergbush \& VandenBerg (1992)} for a metal abundance
${\rm [Fe/H]} = -1.68$ ($Z = 0.0004$).
In Fig.~\ref{tips_age} we compare  the observed V absolute magnitude of the RGB
tip, $M_V^{\rm TRGB}$, with the $V$ luminosity of the RGB tip from the
\bib{Bergbush \& VandenBerg's (1992)} models, for different ages and
metallicities, assuming a distance of 870 pc (see below).
$M_V^{\rm TRGB}$ is a sensitive function of metallicity,
 becoming fainter as the metal content grows, while it is little
sensitive to age for stellar populations older than 8 Gyr
(in contrast, the LF can be a useful age indicator for younger stellar
populations).
If Tucana's dominant population is similar
to old globular clusters (a plausible assumption given the absence of any
significant intermediate-age population),
the location of the observed tip (shaded region in Fig.~\ref{tips_age})
is consistent with models with $Z = 0.0004$\ ([Fe/H] $=-1.68$),
in very good agreement with the metallicity estimates given in 
Sect.~\ref{sec_metal}.

The $I$ magnitude of the tip of the first-ascent red giant branch can be
used to determine the distance to Tucana, following the methods described in
\bib{Mould \& Kristian (1986)} and \bib{Lee \etal (1993a)}.
Distance estimates obtained using the RGB tip have been shown to be
of comparable accuracy to using Cepheid or RR Lyrae variables
(\bib{Madore \& Freedman 1995}).
This method has the considerable advantage that
for old, metal-poor systems ($-2.2 <$ [Fe/H] $< -0.7$,
 $2< t <15$ Gyr), the $I$ absolute magnitude of the RGB tip is almost
independent of abundance
($M_I \simeq -4.0 \pm 0.1$ mag), so that mutually independent distance and
metallicity estimates can be obtained.
Following \bib{Lee \etal (1993a)},
the RGB tip was measured by convolving
the $I$ luminosity function with a zero-sum Sobel kernel: 
the convolved magnitude distribution shows an evident peak
at $I = 20.7$, with an 
estimated uncertainty of the order 0.15 mag.

The distance was then calculated from the $I$ absolute magnitude of the RGB
tip, using the relations of DA90 yielding the bolometric luminosity of the RGB
tip as a function of metallicity, and the bolometric correction as a function
of color:

\begin{eqnarray}
  BC_I &=& 0.881 - 0.243\,(V-I)_{\rm tip} \\
  M_{\rm bol} &=& - 0.19\,{\rm [Fe/H]} - 3.81.
\end{eqnarray}

\smallskip \noindent
Assuming [Fe/H] $= -1.82$ for the metallicity (cf. Sec.~\ref{sec_metal}),
and $(V-I)_{\rm tip}=1.45\pm0.05$ (cf. Fig.~\ref{cmdi}),
we obtain $BC_I = 0.531\pm0.01$,
$M_{\rm bol}^{\rm TRGB} = -3.46\pm0.04$,
$M_I^{\rm TRGB} = -3.99 \pm0.04$, and $M_V^{\rm TRGB} = -2.54 \pm0.04$.
The distance modulus to Tucana is then
$(m-M) = 24.69\pm0.16,$ corresponding to $870 \pm 60$\ kpc.

\section{Metallicity}
\label{sec_metal}

\subsection{Mean abundance}

We have estimated the metallicity of Tucana using two different procedures,
both based on the globular cluster data of \bib{Da Costa \& Armandroff (1990)}.
First, the abundance has been estimated from $(V-I)_{-3.5}$, the color
 of the RGB measured at absolute magnitude $M_I = -3.5$, 
using the relation
\begin{equation}
{\rm [Fe/H]} = -12.64 + 12.6\,(V-I)_{-3.5} - 3.3\,(V-I)_{-3.5}^2
\end{equation}

\smallskip \noindent
(\bib{Lee \etal 1993a}).

$(V-I)_{-3.5}$\ was obtained from a $3\sigma$-clipped Gaussian fit to
the color distribution of the giant stars in the range $21.13 < I < 21.33$.
The distribution is peaked at $(V-I)_{-3.5} = 1.31 \pm 0.03$\ (standard error
of the mean), yielding ${\rm [Fe/H]} = -1.82 \pm 0.12 $.
The reddening towards Tucana is $E_{(B-V)}=0.00\pm0.01$
(\bib{Burstein \& Heiles 1982}), so that
absorption and reddening corrections are negligible.

\begin{figure}
\psfig{figure=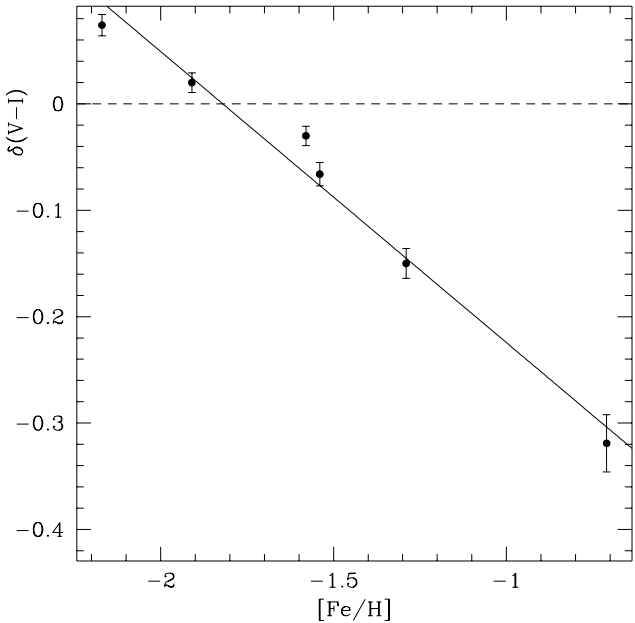,width=8.8cm}
\caption{The mean difference between Tucana's fiducial RGB points and
fiducial RGB
lines from Da Costa \& Armandroff (1990) is plotted against [Fe/H].
The error bars are the standard deviations of the residuals in $(V-I)$.
The solid line represents a linear fit to the points (see text) }
\label{feh_dvi}
\end{figure}

Second, the metallicity of Tucana was determined by direct comparison
of the Tucana's RGB with the globular cluster giant branches of DA90.
The mean color differences $\delta(V-I)_0$ between the RGB sequences
of the 6 GC's in Fig.~\ref{cmdi} and our  Tucana's giant branch have been
calculated using the brightest 1.8 mag of the sequence ($I \leq 22.5$).
The color differences are plotted in Fig.~\ref{feh_dvi} against globular
cluster metallicities.

In order to reduce the statistical noise, we used only stars
in the 2-$\sigma$\  sample.
A linear fit to the mean color differences as a function of GC metallicities
on the DA90 system is

\smallskip\noindent
\begin{equation}
\delta(V-I)_0 = -0.274\,{\rm [Fe/H]} - 0.498.
\label{vi_to_metal}
\end{equation}

\smallskip\noindent
The interpolated metal abundance of Tucana is [Fe/H] $= -1.81$, with
a  dispersion about the regression line
 $\sigma_{(V-I)}$ = 0.01 (standard deviation of the mean). 
Adding quadratically the systematic uncertainty 0.03 mag on the absolute
color calibration, we obtain a total uncertainty
$\sigma[\delta(V-I)] = 0.03$, and $\sigma{\rm [Fe/H]} = 0.11$
dex for the metallicity error. 
This error must be considered as an internal
error, intrinsic to the method we have adopted to estimate the metallicity.
A further error source comes from the uncertainty in the distance 
modulus. Assuming a distance modulus error of 0.16 mag (cf. 
Sec.~\ref{sec_lf}), we have further an uncertainty of $\pm0.13$ on 
[Fe/H].
To these error sources, 
we need to add the intrinsic uncertainty in the GC metallicity,
which is of the order of $\pm 0.1$ dex (Zinn and West 1984), making the total
uncertainty for Tucana's metallicity of 0.20 dex. 

Since stars on the red and asymptotic giant branches
cannot be separated with the present photometric errors ($\leq$ 0.2 mag),
we discuss here a possible bias introduced by the presence of
the old-population AGB.
The brightest AGB stars are found 1.2 and 1.4 mag below the RGB tip in the
$I$ -- $(V-I)$\  color magnitude diagrams of M15 and M2, respectively.
Since the metallicity of Tucana is intermediate between that
of M15 ([Fe/H]=$-2.17$) and M2 ([Fe/H]=$-1.58$), we assume
that the upper 1.3 mag of the Tucana's RGB is unaffected by AGB stars,
under the hypothesis that the bulk of its stellar population
is similar to that of old globular clusters.
Therefore, we have repeated the interpolation using the
brightest 1.2 mag of the giant branch, and selecting only stars with
$(V-I) > 1.15$, and found [Fe/H]$=-1.83$.
We conclude that the average metallicity of Tucana is 
[Fe/H]=--1.82$\pm$0.20
and adopt this value in the present paper.

\subsection{RGB width}

The RGB color spread of several dwarf spheroidals has been found to be larger
than measurement errors, a result generally interpreted as due to an abundance
spread of the red giant population. This is not a general rule, however.
For instance, \bib{Smecker-Hane \etal (1994)} have been able to separate the
contributions of the RGB and the intermediate-age AGB found in Carina,
suggesting that the width of the giant branch is consistent with the RGB of
Galactic globular clusters. 
The presence of a metal-abundance spread has received in some cases
spectroscopic confirmation, \eg from spectrophotometry of individual giants
in Draco (\bib{Lehnert \etal 1992}) and Sextans (\bib{Suntzeff \etal 1993}).
For example, from a model atmosphere analysis of
14 giants in Draco, \bib{Lehnert \etal (1992)}
conclude that the color spread of Draco's giants is due
to a $\sim1$\ dex range in metal abundance. Such a range of chemical
abundance is traced back to chemical enrichment of the interstellar
medium by supernova ejecta and stellar winds.

Here we have measured the RGB width
from our photometry to see if it is compatible with internal errors.
Clearly, tighter upper limits on an intrinsic abundance dispersion
are expected from HST observations.
Tab.~\ref{fiducial} lists the width of the RGB as measured from the
$1\farcs2$ and the $1\farcs5$ image sets. The two sets give
similar results. By comparing the colors 
from the two sets, we have obtained an estimate of the
internal errors which is independent of an AGB contribution to the color
spread.
In a given magnitude bin, we have calculated $\sigma_{\Delta(V-I)}$,
the standard deviation of the difference between the color measurements 
of the same stars in the two image sets.
This reflects the combined measurements errors.
Assuming that the color measurements are of comparable uncertainty
in the two image sets, and uncorrelated, we can estimate
the color measurement error
$\sigma_{\rm meas}  = \sigma_{\Delta(V-I)} / \sqrt{2}$.
The values of $\sigma_{\rm meas}$ are listed in
Table~\ref{fiducial}, 
whereas the $1\sigma$
color errors derived from artificial star experiments on the $1\farcs2$ data
are listed in  Table~\ref{c_and_err}.
While, in principle, crowding experiments (which use a constant PSF) could
underestimate the random measurement errors in the case of large
PSF variations across the frame, in the present case
random errors estimated from simulations and from
comparison of different frames are perfectly consistent.
Note that the low value of $\sigma_{\rm meas}$ for the faintest bin
is an artifact of incompleteness in the $1\farcs5$ frames.

The results in Tab.~\ref{fiducial} and Tab.~\ref{c_and_err} show that
the width of the Tucana's giant branch is not significantly different 
from that expected on the basis of photometric errors. 
Once measurement errors are critically and carefully assessed,
no convincing indication of an abundance spread is found in our data.

\section{Surface photometry and structural parameters}
\label{sec_surfot}

\subsection{Surface photometry}
\label{subsec_surfot}

In this section we derive the structural parameters of Tucana needed
to compare this galaxy with the other LG dwarf spheroidals.
Our procedure follows quite closely the methods of \bib{Caldwell \etal (1992)}.
In order to measure the luminosity profile of Tucana,
all stars brighter than $I=20$ or redder than
$V-I = 1.84$ were subtracted from the coadded images using DAOPHOT.
Further, all bright stars, obvious background galaxies, and diffraction
spikes  were interactively masked out.
In this way we were left with stars mostly belonging to
the RGB, superimposed onto a diffuse, unresolved light component.
Surface photometry was performed only on the $V$\ image, since flat-fielding of
the outermost area of the $I$\ image was not of sufficient accuracy to 
derive an extended surface brightness profile.
The cleaned master $V$\  image was processed as follows. After  $2\times2$
rebinning, the frame was median filtered with a $5\times 5$ box.
The smoothed image thus produced is essentially elliptical, with intrinsic
noise due to the presence of resolved stars with a distribution which appears
somewhat clumpy.
Ellipses were fit to isophotes in the radial range
$37\arcsec < a < 138\arcsec$
($a$ being the semi-major axis), leaving all parameters free.
Given the noise on the parameters of each ellipse,
we restrict ourselves to providing mean values of ellipticity
($\epsilon =  1 - b/a$) and position angle (PA) for $a > 65\arcsec$.
These are given in Table~\ref{struct_par} along with their standard errors.

\begin{figure}
\psfig{figure=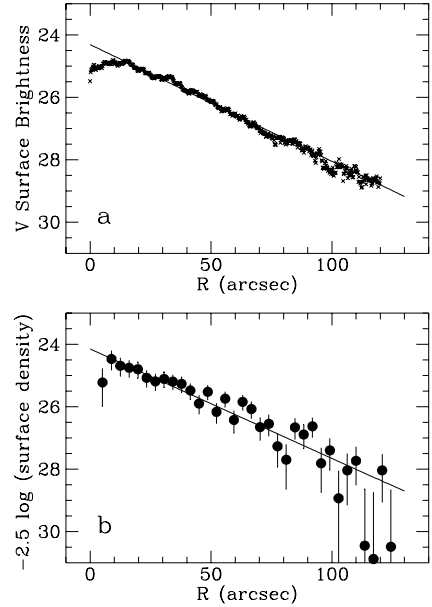,width=8.8cm}
\caption{{\bf a}\  Surface brightness in $V$ (mag arcsec$^{-2}$) against
the geometric radius for Tucana. The line represents an exponential
fit to the luminosity profile between $10\arcsec$ and $100\arcsec$,
with scale length $29\arcsec$.
{\bf b} Surface density (arcmin$^{-2}$) of resolved stars in the
Tucana field. Abscissa is geometric radius. Also shown is an exponential
fit with scale length $31\arcsec$ (solid line)
}
\label{surf_newfig2}
\end{figure}

A $V$ surface brightness profile including galaxy $+$ sky was then obtained
as the median flux along ellipses with constant ellipticity and orientation,
and plotted against geometric radius ($r_{\rm g} = \sqrt{ab}$).
The azimuthally averaged profile is essentially flat for $r_{\rm g} >
140\arcsec$, so that the mean intensity 
for $r_{\rm g} > 140\arcsec$\ was chosen to represent the sky level.
In this range, the rms scatter of the profile, binned in 10\arcsec\
intervals, is $\sim 0.05$\%.
Large-scale ($> 20\arcsec$) peak-to-valley
fluctuations of the sky background in the master $V$\ frame,
in the absence of azimuthal averaging, are of the order 0.3\%.
By extrapolating the inner ($r_{\rm g} < 100\arcsec$)
exponential profile out to this
geometric radius, we estimate that the galaxy light at $r_{\rm g} =
140\arcsec$\ is of the order 0.1\% of  (or about 7.5 mag arcsec$^{-2}$ below)
the sky background. This is comparable to the uncertainties on
the sky level, and
the profile can be reliably measured down to $\mu_V \sim 29$ mag
arcsec$^{-2}$, where the galaxy light contribution becomes comparable to
$3\sigma$\ sky fluctuations.
The sky-subtracted $V$ luminosity profile of Tucana is shown in
Fig.~\ref{surf_newfig2}$a$  against geometric radius, along with
an exponential fit in the range $10\arcsec-100\arcsec$.
As it is common for dwarf elliptical galaxies
(\cf \bib{Caldwell \etal 1992}), the surface brightness profile of Tucana
is well fit by an exponential law except for the center.

\tablecaption{Structural parameters for Tucana }
\tablelabel{struct_par}
\begin{planotable}{lrcl}
\tablehead{
\colhead{Parameter~~~~~~~} &
\colhead{} &
\colhead{Value} &
\colhead{} }
\startdata
 $V_T$                 & 15.15 & $\pm$ &  0.18 \\
 $M_V$                 & $-9.55$ & $\pm$ &  0.27 \\
 $\mu_{0,V}$ {(obs)}   & 25.05 & $\pm$ &  0.06  \\
 $\alpha$ (pc)         & 123 & $\pm$ &  19 \\
 $S_{0,V}$             & 24.30 & $\pm$ &  0.19 \\
 $R_e$\ (pc)           & 205 & $\pm$ &  32 \\
 $SB_{\rm e}$          & 25.43 & $\pm$ &  0.19  \\
 $r_{\rm c}$ (pc)      & 176 & $\pm$ &  26  \\
 $r_{\rm s}$ (pc)      & 166 & $\pm$ &  5  \\
 $\mu_{0,V}$ (King)    & 24.76 & $\pm$ &  0.18 \\
 $c$                   & 0.72 & $\pm$ &  0.12 \\
 $1-b/a$               & 0.48 & $\pm$ &  0.03  \\
 P.A.                  & $97^\circ$  & $\pm$ &   $2^\circ$  \\
\end{planotable}

The scale length ($\alpha$) and central extrapolated surface brightness of
the fit ($S_{0,V}$) are specified in Table~\ref{struct_par}, 
assuming a distance of 870 kpc.
Note that all magnitude errors given in Table~\ref{struct_par}
reflect internal error estimates; a 0.02 mag zero point uncertainty should
be added to account for systematic errors.
Our best fit gives a scale length of $29\arcsec\pm5\arcsec$, or
123 pc.
This scale length is consistent with that found by
\bib{Da Costa (1994)}, and makes Tucana one of the
smallest dwarf spheroidals known.
Our extrapolated central surface brightness
$S_{0,V} =24.3$\  mag arcsec$^{-2}$ is
somewhat brighter than that measured by Da Costa
(24.8 mag arcsec$^{-2}$ assuming a central color $B-V = 0.7$).
The uncertainties given in Table~\ref{struct_par} are the maximum errors 
obtained by changing the sky level by $\pm$1\%, 
three times larger than  the sky fluctuations. These are larger than the 
internal errors of the fit.

The total magnitude of Tucana ($V_T=15.15\pm0.18$) was calculated by
integrating the observed surface brightness profile
in the innermost region ($0\arcsec < r_{\rm g} < 20\arcsec$), and
adding a correction based on the exponential fit.
Alternatively, $V_{\rm T}$\  was obtained from the observed profile alone,
integrated out to $r_{\rm g} = 150\arcsec$, in which case
the asymptotic value is $V_T = 15.05$. The two methods give consistent 
results within the quoted errors.
The absolute magnitude $M_V = -9.6$ mag is consistent, within the large
uncertainties inherent in this kind of measurements for dwarf spheroidals,
with the values derived by previous authors.
The effective geometric radius $R_e$, as well as $SB_e$, the mean surface
brightness inside $R_e$, were  derived from the parameters of the
exponential fit.
The errors on all derived quantities were obtained by assigning
their extreme values to $S_{0,V}$\  and  $\alpha$. 
In the case of $M_V$\ we have also taken into account a 0.2 mag error
on the distance modulus.

A King law is also well fit to the Tucana surface brightness profile
(we used in this case the same radial range as for the exponential fit).
A fit was obtained allowing all
parameters to vary. This yielded the central surface brightness of
the model profile, $\mu_{0,V} = 24.76$ mag arcsec$^{-2}$\
(not to be confused with the intensity scale parameter of the King law),
the ``core radius'' $r_s$, and the concentration parameter $c$,
all  listed in Table~\ref{struct_par}.
For comparison, the core radius at which the surface brightness drops at half
the central value is $r_{\rm c}$ = 42\arcsec $\pm$ 6\arcsec\  ($176\pm26$\ pc).
Uncertainties on these parameters were estimated by changing the 
sky level by $\pm3\sigma$, and the range over which the fit was made. 

Also in the case of the King fit, the central surface brightness of
the fit profile, $\mu_{0,V} {\rm (King)}$, is higher than the observed value,
$\mu_{0,V} {\rm (obs)}= 25.05 \pm 0.06$ mag arcsec$^{-2}$.
The latter was calculated as the median surface brightness in the inner
3\arcsec, together with its associated rms scatter.
The dip is certainly real, since it is also seen in the $I$ profile, and
the $B$ profile of \bib{Da~Costa (1994)}. Inspection of the images seems to
suggest a clumpy (``ring-like'') distribution of the bright giants in the
central region, rather than evidence for a dust lane. 

\begin{figure}
\psfig{figure=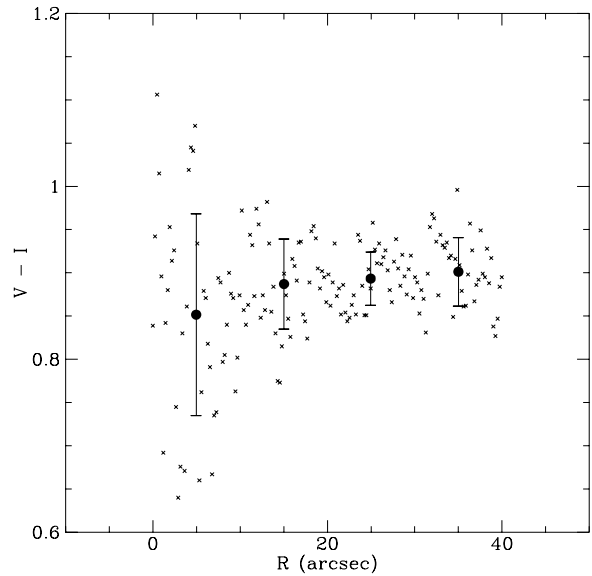,width=8.8cm}
\caption{Median $(V-I)$ color averaged on ellipses plotted
against distance from the galaxy center. Dots
with error bars represent mean values in $10\arcsec$ bins}
\label{surf_fig1}
\end{figure}

Finally, Fig.~\ref{surf_fig1} shows the integrated color of
Tucana inside $r_{\rm g}=40\arcsec$.
The central $(V-I) = 0.89\pm0.01$ was obtained as the mean of the inner
$20\arcsec$\   ($V-I = 0.87$\ inside 40\arcsec).
Adopting the mean relation $(B-V) = 0.8\,(V-I)$\  derived for the
giant branches of globular clusters 
(\bib{see Smecker-Hane \etal 1994}), this value
corresponds to $(B-V) = 0.70$,
in full agreement with the result of \bib{Da Costa (1994)}.

\subsection {Surface density of resolved stars}

We have also investigated  the structure of Tucana using
the surface density of stars as a function of distance from the
galaxy center.
Figure~\ref{surf_newfig2}$b$  shows the surface density
profile obtained by counting
stars in elliptical annuli with a $5\arcsec$ step in semi-major axis,
and fixed ellipticity and position angle.
Only stars in the magnitude range  $20.7 < I < 22.5$ have been considered.
The counts have been corrected for the radially varying incompleteness.
The background+foreground counts have been estimated from the
count level beyond $200\arcsec$ to be $7.6 \pm 1.0$ arcmin$^{-2}$
(standard deviation refers to the mean value of 29 bins).
The small number of stars inside a $10\arcsec$\ radius is only partially
explained by incompleteness.
An exponential fit to the counts between 20\arcsec\ and 90\arcsec\  yields
a scale length $\alpha = 31\arcsec \pm 3\arcsec$\  ($131\pm13$  pc), in good
agreement
with the results from surface photometry.
The uncertainty has been estimated by empirically varying
the background value by $\pm 5\sigma$\
and repeating the fit in the same radial interval.

\subsection {Absolute magnitude from the LF}

The completeness-corrected $V$ luminosity function
has also been used to estimate the absolute magnitude
$M_V$ of Tucana by comparison with the LF of the globular
cluster M3.
The corrected $V$ luminosity function was 
scaled by a factor 1.20 to match the luminosity function of  M3
from \bib{Sandage (1954)}.
The logarithmic shift is $0.083\pm0.055$ (the error being the
standard deviation of the differences on a logarithmic scale).

Since the magnitude
distribution in Fig.~\ref{lf} refers to the inner region,
an extrapolation factor to total counts over the entire galaxy  was
obtained by integrating the surface brightness profile.
Assuming that the luminosity profile of Tucana follows
an exponential law with a 30\arcsec\  scale length, we find
that the inner region contains $\sim$73\% of the Tucana stars.
Further, we assume that 5\% of the RGB
stars are left out from the $2\sigma$-sample.
Therefore the number of stars in the Tucana's LF should be multiplied
by 1.45, so that  $L_{\rm Tuc}=L_{\rm M3}*1.75$.
Assuming $M_V^{\rm M3} = -8.75$\ (\bib{Djorgovski 1993}), we obtain
for Tucana $M_V = -9.4 \pm 0.28$, in very good agreement with the
absolute magnitude estimate given in Sect.~\ref{subsec_surfot}.
The errors are estimated taking into
account the rms uncertainty in the sliding fit of the LF's,
a 10\% uncertainty on the scale length, and the uncertainty
on the distance modulus.

\section{Discussion}
\label{sec_discu}

Deep color-magnitude diagrams obtained in the last decade have revealed
a wide variety of star formation histories in dwarf spheroidals, and
direct evidence for recent star formation has been found in 50\% of the
dE/dSph galaxies in the LG (\bib{Freedman 1994}).
A complex star formation history seems also to be fairly common in Virgo 
and Fornax dE's (see \bib{Ferguson \& Binggeli 1994}; 
\bib{Held \& Mould 1994};  and references therein).
The presence of carbon stars in many dSph companions to the Galaxy,
 and in And~II, 
(see \bib{Armandroff 1994} for a recent review) 
indicates that star formation took place
over an extended time period, probably in episodes alternated to
quiescent phases.

A possible correlation between the star formation history of dSph and their
distance from the parent galaxy was noted by \bib{van den Bergh
(1994)}, in the sense that young or intermediate-age populations are 
preferentially found  in dSph's far from the Milky Way. 
He suggested that 
dSph close to the Galaxy formed the bulk of their stars at early epochs,
either because efficient star formation was triggered by interaction,
or because any remaining gas was efficiently removed
by supernova-driven winds or high UV flux. 
According to this trend, a distant dwarf such as Tucana would be expected to
harbour an intermediate-age population, witnessing a continued history of star
formation, as in Carina (\eg \bib{Smecker-Hane \etal 1994}) and 
Leo~I (\bib{Lee \etal 1993b}). 

Our study confirms that Tucana is located in a 
no--man's--land, close to the border of the LG, at 870 Kpc from
the Galaxy and $\sim 1300$ Kpc from both M31 and M33. There are 
other isolated dwarf galaxies in the LG (\eg Phoenix, WLM, NGC 
6822, DDO 210, etc...): all of them have experienced recent star 
formation. 
As suggested by \bib{Da Costa (1994)} and shown 
in Sec.~\ref{sec_cmd}, there is no sign of intermediate or young population
in the color--magnitude diagram of Tucana, thus 
Tucana does not follow the proposed trend of the star formation 
history in the LG galaxies. 
This  absence of an intermediate-age population seems to imply that
galactic winds expelled the interstellar medium at early epochs and halted 
star formation, a process apparently not dependent on environment. 
One further question rises immediately: is Tucana an unique example or
are there many other dwarf spheroidals located far from the largest
galaxies in the  Local Group ? 
Clearly, the discovery of dwarf galaxies located in the 
outer parts of the LG is biased in favour of star forming 
objects. On the other hand, at distances beyond 400--500 Kpc and out to the 
borders of the LG ($\sim 1$ Mpc), identifying new dSph's like Tucana 
is extremely difficult with the current survey techniques
(\bib{Irwin 1994}), and the search must be considered highly incomplete.

\begin{figure}
\psfig{figure=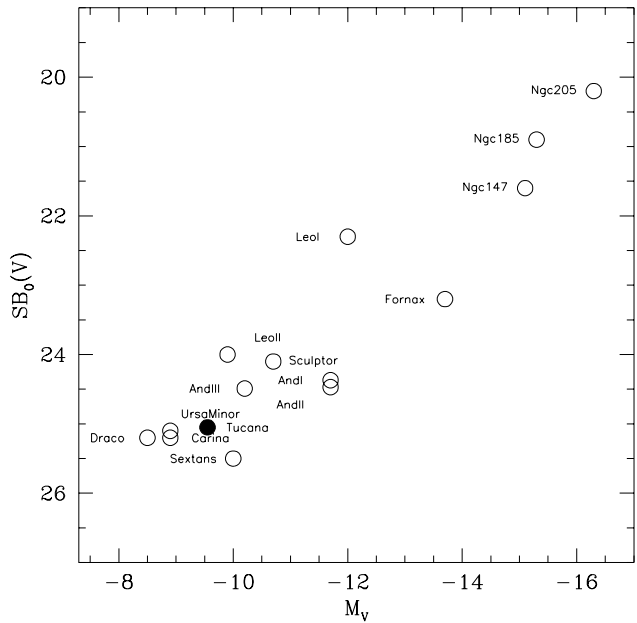,width=8.8cm}
\caption{Central surface brightness vs. absolute $V$ magnitude for LG
dwarf spheroidals}
\label{sv0}
\end{figure}

\begin{figure}
\psfig{figure=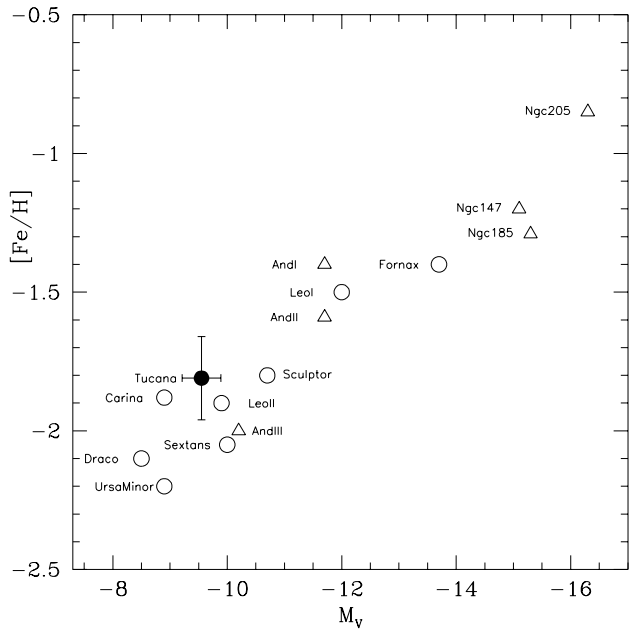,width=8.8cm}
\caption{The metallicity--luminosity relation defined by the dSph companions of
the  Galaxy ({open circles}) and of M31 ({open triangles}). Tucana
({full dot}) obeys to the same relation}
\label{mv_feh}
\end{figure}

While Tucana is peculiar as far as its star formation history is 
concerned, in many other respects it is indistinguishable from other 
LG dSph's or dE's. It does not contain large amounts of HI,
though the $3 \sigma$ non--detection upper limit of $M_{HI}/L_B<2.5$ 
(Lavery and Mighell 1992) 
still corresponds to $\sim 10^6$\  M$_\odot$, and a better 
estimate of the HI content would be needed. Tucana's 
morphological parameters are within the range of the Galaxy and M31 dSph 
companions, and it conforms to the surface brightness--absolute 
magnitude--metallicity relations defined by the LG dSph
(\cf \bib{Da Costa 1994}). 
In Fig.~\ref{sv0} we have plotted the central surface brightness of 
dSph's of the LG against their absolute
magnitude. Tucana clearly lies on the single relation defined by
the Milky Way and M31 companions (\bib{Caldwell \etal 1992}).
The surface brightness seems to be an intrinsic property of dSph's,
not related to proximity to a bigger galaxy. Since surface density
probably reflects the evolution of dwarfs through mass loss, the implication
would be that gas loss is more dependent on intrinsic galaxy properties
(\eg mass) than on interaction with environment.

\bib{Bellazzini \etal (1996)} have recently proposed that
the central surface brightnesses of the dwarf spheroidal satellites
of the Milky Way are correlated
with their Galactocentric distance, in the sense that
dwarfs closer to the Galactic Center are dimmer. An even tighter
correlation is found between surface brightness and a
linear combination of distance and absolute magnitude, which seems to
hold also for the M31 dSph satellites.
This may imply that their structure is influenced by the environment.
Having a ``normal'' central surface brightness for its absolute
luminosity, and a large galactocentric distance, Tucana
clearly deviates from their proposed correlation. 
It is not clear whether the \bib{Bellazzini's \etal (1996)} trends mainly
reflect a correlation between $R_{\rm GC}$ and luminosity. 
In addition, it should be borne in mind that selection effects may be 
operating against the discovery of dim dwarfs at large distances. 

This study confirms that Tucana has a low metallicity: this distant, isolated
dwarf follows (Fig.~\ref{mv_feh}) 
the metallicity-luminosity relation defined by the M31 and Milky
Way dSph companions (\bib{Caldwell \etal 1992}). 
This is a further evidence that formation and evolution of the dSph's is
not strongly influenced by the environment (\cf \bib{Armandroff 1994} and
references therein). 
As the metallicity appears to be a function of luminosity only, the
metal enrichment process itself seems not to be influenced by the presence of
a nearby galaxy (\bib{Da Costa 1994}). 
Does the metallicity of Tucana imply self-enrichment with  some retention
and/or re-capture of gas (\bib{Silk, Wyse, \& Shields 1987}) ?
Because of the isolated location of Tucana and the absence of an
intermediate-age population, re-capture of gas as an effect of interaction
seems unlikely as it is unlikely that the metallicity 
of this dwarf be influenced by accretion of pre-enriched external 
material.  Thus it may represent an ideal laboratory for studying chemical 
evolution of dwarf spheroidals. 

As a final comment, we note that measuring the radial velocity of Tucana, in
addition to the accurate distance given here,  will allow to use this isolated
galaxy as a test particle to constrain the mass distribution and dynamics of
the Local Group (\bib{Zaritsky 1994}). 
Moreover, obtaining high-precision velocities for the brightest giants in
Tucana, to measure the internal velocity dispersion of Tucana and infer its
dark matter content, would be a valuable task for a 10-metre class telescope.

\acknowledgements{
I.S. acknowledges support by MURST during his
``Dottorato di Ricerca'' fellowship. 
We thank Dr. J. Bahcall for providing the computer code of his models, 
Dr. G. Clementini for sharing her extinction measurements, and 
the  La Silla observatory staff for support. 
E.H. wishes to thank Jeremy Mould for useful conversations regarding 
dwarf galaxies. 
Finally, we wish to thank the referee, Dr. G. Da Costa, for suggestions
that improved the presentation of this paper.
}

\enddocument